\documentclass[conference]{IEEEtran}
\IEEEoverridecommandlockouts

\usepackage{graphicx}
\usepackage{amsmath, amssymb, amsfonts}
\usepackage{listings}
\usepackage{xcolor}
\usepackage{hyperref}
\usepackage{balance}
\usepackage{caption}
\usepackage{subcaption}
\usepackage{algorithm}
\usepackage{algorithmic}
\usepackage{tikz}
\usetikzlibrary{arrows.meta, positioning}

\definecolor{codegray}{rgb}{0.5,0.5,0.5}
\definecolor{codegreen}{rgb}{0,0.6,0}
\definecolor{codepurple}{rgb}{0.58,0,0.82}
\definecolor{backcolour}{rgb}{0.95,0.95,0.92}

\lstdefinestyle{mystyle}{
    backgroundcolor=\color{backcolour},   
    commentstyle=\color{codegreen},
    keywordstyle=\color{blue},
    numberstyle=\tiny\color{codegray},
    stringstyle=\color{codepurple},
    basicstyle=\ttfamily\footnotesize,
    breakatwhitespace=false,         
    breaklines=true,                 
    captionpos=b,                    
    keepspaces=true,                 
    numbers=left,                    
    numbersep=5pt,                  
    showspaces=false,                
    showstringspaces=false,
    showtabs=false,                  
    tabsize=2
}

\begin{document}

\title{Bugdar: AI-Augmented Secure Code Review for GitHub Pull Requests}

\author{\IEEEauthorblockN{John E. Naulty}
\IEEEauthorblockA{
\textit{Mysten Labs}\\
Palo Alto, CA, USA \\
jnaulty@mystenlabs.com}
\and
\IEEEauthorblockN{Eason Chen}
\IEEEauthorblockA{
\textit{Carnegie Mellon University}\\
Pittsburgh, PA, USA \\
EasonC13@cmu.edu}
\and
\IEEEauthorblockN{Joy Wang}
\IEEEauthorblockA{
\textit{Mysten Labs}\\
Palo Alto, CA, USA \\
joy@mystenlabs.com}
\and
\IEEEauthorblockN{George Digkas}
\IEEEauthorblockA{
\textit{Mysten Labs}\\
Palo Alto, CA, USA \\
george.digkas@mystenlabs.com}
\and
\IEEEauthorblockN{Kostas Chalkias}
\IEEEauthorblockA{
\textit{Mysten Labs}\\
Palo Alto, CA, USA \\
kostas@mystenlabs.com}
}
\maketitle

\begin{abstract}
As software systems grow increasingly complex, ensuring security during development poses significant challenges. Traditional manual code audits are often expensive, time-intensive, and ill-suited for fast-paced workflows, while automated tools frequently suffer from high false-positive rates, limiting their reliability. To address these issues, we introduce Bugdar, an AI-augmented code review system that integrates seamlessly into GitHub pull requests, providing near real-time, context-aware vulnerability analysis. Bugdar leverages fine-tunable Large Language Models (LLMs) and Retrieval Augmented Generation (RAGs) to deliver project-specific, actionable feedback that aligns with each codebase's unique requirements and developer practices. Supporting multiple programming languages, including Solidity, Move, Rust, and Python, Bugdar demonstrates exceptional efficiency, processing an average of 56.4 seconds per pull request or 30 lines of code per second. This is significantly faster than manual reviews, which could take hours per pull request. By facilitating a proactive approach to secure coding, Bugdar reduces the reliance on manual reviews, accelerates development cycles, and enhances the security posture of software systems without compromising productivity.
\end{abstract}


\section{Introduction}

Modern software systems have become increasingly complex, creating substantial challenges for security auditing processes. For example, in the DeFi sector, manual audits often take weeks to complete, with costs exceeding tens of thousands of dollars per project \cite{cyberscope2022audit}. These manual audits, while thorough, are slow and expensive, delaying critical deployments. Automated tools offer speed but lack precision, often generating excessive false positives \cite{6606613}. This trade-off highlights the need for an innovative solution that combines human-level expertise with the efficiency of AI.

To address these challenges, we present Bugdar, an AI-augmented code review system that transforms the GitHub pull request workflow into a secure, efficient, and developer-friendly environment. Bugdar integrates fine-tunable Large Language Models (LLMs) to identify potential security flaws. Unlike traditional tools, Bugdar adapts dynamically to project-specific contexts by employing Retrieval Augmented Generation (RAGs), enabling context-aware and actionable feedback tailored to the unique characteristics of decentralized applications, smart contracts, and multi-language codebases.

To be specific, Bugdar incorporates various features to enhance secure coding practices and streamline the development workflow. These features include:

\begin{itemize}
    \item \textbf{Structured Vulnerability Reports:} Generates detailed reports outlining potential security vulnerabilities, including severity, potential impact, and recommended remediation steps, mimicking the comprehensive output of experienced human auditors.
    \item \textbf{Prompt Engineering and Fine-tuning:} Allows developers to customize the prompts used by the LLMs to target specific security concerns and fine-tune the models using project-specific datasets and human feedback.
    \item \textbf{Multi-Programming-Language Support:} Analyze code written in various programming languages, including Move and Solidity smart contract languages, as well as Rust, TypeScript, Python, and Go, enabling a comprehensive security assessment across diverse technology stacks.
    \item \textbf{GitHub CI/CD Integration:} Integrates as a GitHub app, providing security analysis and feedback directly within the pull request interface.
\end{itemize}

These features collectively transform the way developers approach secure coding practices. By integrating seamlessly into the GitHub workflow, Bugdar fosters a proactive security mindset, enabling developers to address vulnerabilities early in the development lifecycle. This approach reduces the reliance on extensive manual reviews, minimizes delays in deployment, and enhances collaboration between security experts and development teams. Bugdar's tailored and context-aware analysis empowers teams to focus on building robust and secure applications without compromising speed or agility.

\section{Background and Related Work}

\begin{figure*}
    \centering
    \includegraphics[width=1\linewidth]{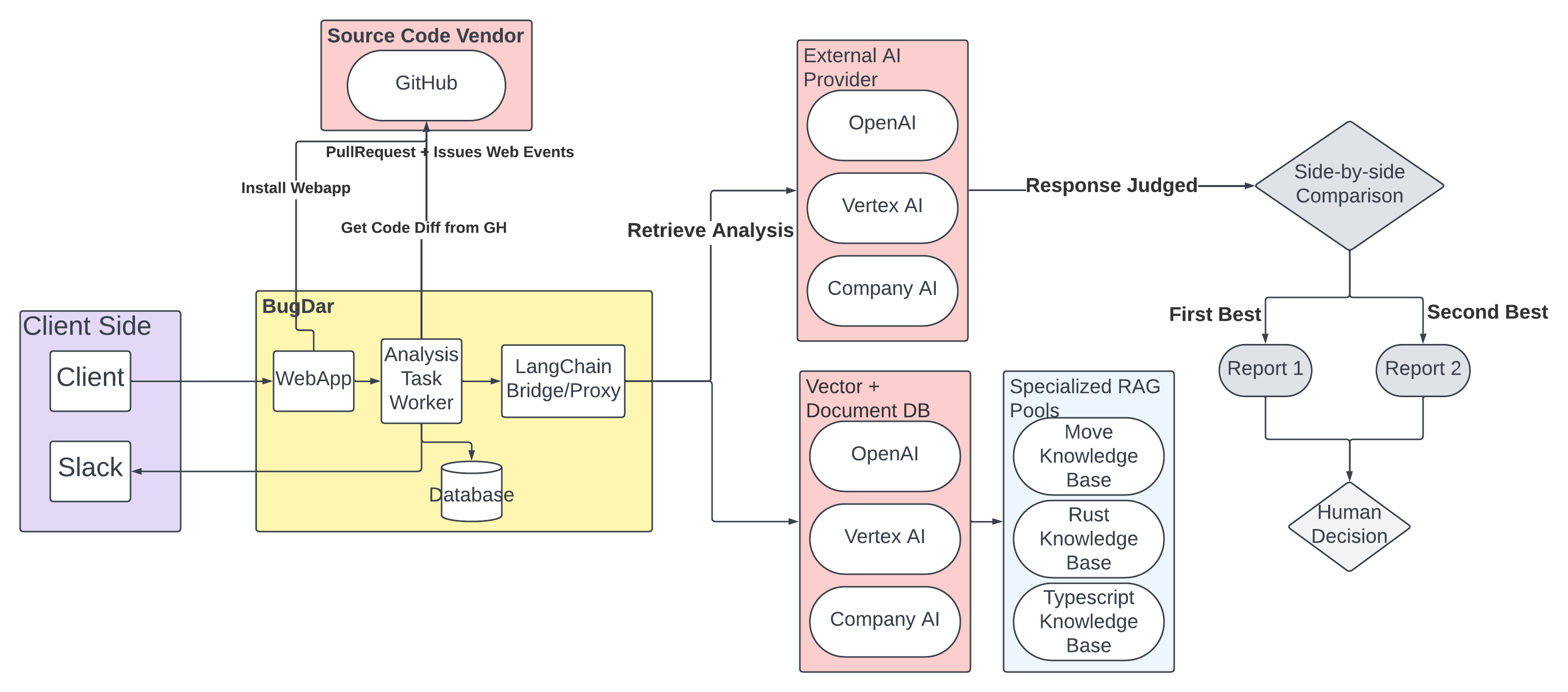}
    \caption{The diagram illustrated the architecture of the Bugdar system.}
    \label{fig:architecture_diagram}
\end{figure*}
Several researchers have investigated the use of LLMs for code-related tasks, including code generation, code summarization, and bug detection\cite{david2023manual,gao2023far,li2021drawbacks, chen2024mad, chen2021evaluatinglargelanguagemodels, chen2023gptutor}. They find that LLMs demonstrated impressive capabilities in understanding and generating code by learning from large datasets of publicly available source code \cite{achiam2023gpt}.

Recent studies have specifically examined LLMs' effectiveness in vulnerability detection. \cite{david2023manual} evaluated GPT-4 and Claude models on DeFi smart contracts, finding they correctly identify vulnerability types in 40\% of cases, though with notable false positive rates. Similarly, VulBench\cite{gao2023far}, a comprehensive vulnerability benchmark, demonstrating that several LLMs outperform traditional deep learning approaches in vulnerability detection.

Traditional machine learning approaches for vulnerability detection often rely on handcrafted features or require structured representations of code, such as abstract syntax trees (ASTs) \cite{li2018vuldeepecker}. However, these methods can be limited in scalability and effectiveness when dealing with complex or obfuscated code \cite{ragkhitwetsagul2019survey}.

Bugdar addresses this challenge by combining the analytical capabilities of LLMs with techniques like Retrieval Augmented Generation (RAGs) to provide context-aware and project-specific security analysis \cite{lewis2020retrievalaugmented}. RAGs enable Bugdar to access and utilize relevant project documentation, including code comments, design specifications, and security guidelines, enhancing its understanding of the codebase and improving the accuracy of its vulnerability assessments.

Bugdar's "Specialization-as-a-Service" model further distinguishes it from existing approaches. This concept emphasizes Bugdar's ability to adapt to the unique characteristics of each project, learning from project-specific data and developer feedback to provide tailored security analysis that aligns with the project's specific context and requirements.

\section{System Development}

Bugdar's architecture is designed to integrate seamlessly into the GitHub pull request workflow, providing developers with timely feedback and actionable security insights.

\subsection{System Architecture}

The architecture is illustrated in \autoref{fig:architecture_diagram}, which includes:

\begin{itemize}
    \item \textbf{GitHub Integration Layer:} Interfaces with GitHub's API to monitor pull requests, retrieve code diffs, and post review comments.
    \item \textbf{Preprocessing Module:} Processes code diffs to prepare input for the LLM, handling tasks such as code chunking and context window management.
    \item \textbf{Context Retrieval Engine:} Utilizes RAGs to fetch relevant project documentation and historical code data.
    \item \textbf{LLM-Based Analysis Engine:} Leverages fine-tunable LLMs (e.g., GPT-4o) to perform security analysis on the code changes.
    \item \textbf{Reporting Module:} Generates structured vulnerability reports and code review comments based on the analysis.
\end{itemize}

\subsection{Analysis Workflow}

The workflow follows the steps in Algorithm \ref{alg:security_analysis}. 
Bugdar's AI-augmented approach aims to enhance secure code review workflow with the following principles:

\begin{algorithm}[h]
\caption{Bugdar Pull Request Analysis Workflow}
\begin{algorithmic}[1]
\REQUIRE New or updated pull request $PR$
\STATE \textbf{Trigger Analysis} upon detection of $PR$.
\STATE \textbf{Fetch Code Diffs} using the GitHub API.
\STATE \textbf{Partition Code Diffs} to fit LLM context window.
\STATE \textbf{Initialize Analysis Report}.
\FOR{each code chunk}
    \STATE \textbf{Analyze Chunk} with LLM(s).
    \STATE \textbf{Select Best Analysis} using a judge LLM.
    \STATE \textbf{Accumulate Results} into the report.
\ENDFOR
\STATE \textbf{Aggregate Results} into a comprehensive report.
\STATE \textbf{Update Credits} based on token usage.
\STATE \textbf{Generate Reports and Notifications}:
    \begin{itemize}
        \item Store the analysis report in the database.
        \item Post summary to the pull request (if enabled).
        \item Send notifications (e.g., Slack) if configured.
    \end{itemize}
\end{algorithmic}
\label{alg:security_analysis}
\end{algorithm}

\begin{itemize}
    \item \textbf{Shifting security analysis earlier in the development cycle:} Providing security feedback during the pull request stage enables developers to identify and address vulnerabilities before they are merged into the main codebase, reducing the cost and effort of remediation later in the development process.
    \item \textbf{Improving developer efficiency:} Automating the analysis process and generating actionable feedback reduces the manual effort required for code review, allowing developers to focus on core development tasks.
    \item \textbf{Fostering a security-conscious development culture:} By integrating security analysis seamlessly into the developer workflow, Bugdar encourages developers to prioritize security considerations from the outset, leading to more secure and robust software systems.
\end{itemize}

\subsection{Dataset}


A comprehensive dataset of real-world GitHub pull requests and source code with known security vulnerabilities was used to evaluate Bugdar. The dataset encompasses a diverse range of projects and programming languages, including Move \cite{blackshear2019move}, Solidity, Rust, TypeScript, and Python, particularly focusing on smart contracts and blockchain-related code. The ground truth for vulnerability classification was established through:

\begin{itemize}
    \item \textbf{Manual Audits by Security Experts:} Ensuring the accuracy of vulnerability labels.
    \item \textbf{Bug Bounty Reports:} Incorporation of real-world security issues from bug bounty programs.
\end{itemize}

\subsection{Evaluation Metrics:}

Bugdar was evaluated based on its ability to classify and describe vulnerabilities identified in the dataset source code. Specifically, the evaluation focused on three key aspects: \textbf{precision, recall, and F1 score} to measure the accuracy of vulnerability detection; \textbf{false positive and false negative rates} to assess the reliability of the analysis; and \textbf{time consumption} to evaluate its efficiency compared to manual reviews and static analysis tools.




\section{Results}

\subsection{Quantitative Results}

\begin{table}[h!]
\centering
\small
\setlength{\tabcolsep}{4pt}
\begin{tabular}{|l|l|c|c|c|c|c|}
\hline
\textbf{Model} & \textbf{Task} & \textbf{RAG} & \textbf{Precision} & \textbf{Recall} & \textbf{F1} & \textbf{Accuracy} \\
\hline
o1-preview & Classification & No  & 0.24 & 0.40 & 0.30 & 0.17 \\
o1-preview & Description    & No  & 0.50 & 0.67 & 0.57 & 0.40 \\
o1-preview & Classification & Yes & 0.43 & 0.43 & 0.43 & 0.27 \\
o1-preview & Description    & Yes & 0.57 & 0.50 & 0.53 & 0.36 \\
\hline
gpt-4o  & Classification & No  & 0.35 & 0.60 & 0.44 & 0.29 \\
gpt-4o  & Description    & No  & 0.58 & 0.73 & 0.65 & 0.48 \\
gpt-4o  & Classification & Yes & 0.39 & 0.64 & 0.49 & 0.32 \\
gpt-4o  & Description    & Yes & 0.43 & 0.59 & 0.50 & 0.33 \\
\hline
\end{tabular}
\caption{Merged Vulnerability Performance (Classification vs.\ Description) With/Without RAG}
\label{tab:merged_vuln_performance}
\end{table}

The evaluation of the Vulnerability Classification and Description is summarized in Table~\ref{tab:merged_vuln_performance}.
In vulnerability classification without RAG, GPT-4o achieved the best performance with a precision of 35\% and recall of 60\%, resulting in an F1 score of 0.44. This indicates that while the model could detect 60\% of existing vulnerabilities, it had a relatively high false positive rate. The o1-preview model showed lower performance with a precision of 24\%, recall of 40\%, and F1 score of 0.30. When using RAG, both models showed improved performance: GPT-4o's metrics increased to 39\% precision, 64\% recall, and an F1 score of 0.49, while o1-preview achieved 43\% precision, 43\% recall, and an F1 score of 0.43.

For vulnerability description tasks, both models showed improved performance. Without RAG, GPT-4o demonstrated stronger results with a precision of 58\%, a recall of 73\%, and an F1 score of 0.65. The o1-preview model also performed better in description tasks, achieving a precision of 50\%, recall of 67\%, and an F1 score of 0.57. With RAG, performance varied: GPT-4o showed precision of 43\%, recall of 59\% and an F1 score of 0.50, while o1-preview achieved precision of 57\%, recall of 50\% and an F1 score of 0.53. These results suggest that while the models are generally more adept at describing potential vulnerabilities than classifying them, the addition of RAG improved classification performance but had mixed effects on description tasks.

\subsection{Time Spent}

The performance of Bugdar in analyzing pull requests was evaluated based on its efficiency compared to human reviews. In total, Bugdar analyzed 14 pull requests, encompassing 23,010 lines of code added, 634 lines removed, and 23,644 lines changed, with an average of 1,643.57 lines added, 45.29 lines removed, and 1,688.86 lines changed per pull request. It processed these 23,644 lines of code in 790 seconds, achieving an average speed of approximately \textbf{30 lines of code per second} or \textbf{56.4 seconds per pull request}. In contrast, manual reviews typically take several hours per pull request, depending on the complexity and size of the code changes. For example, according to our internal data, a security audit of a new future consisting of 10,000 lines of Move smart contract code took two security engineers 10 days--\textbf{0.11 lines of code per second}, and not all security vulnerabilities were identified. This demonstrates that Bugdar achieves a significant reduction in review time, processing code at least \textbf{100x faster} than human reviewers. This time efficiency positions Bugdar as a valuable tool for accelerating secure code reviews, particularly for large-scale or high-frequency development workflows.

\subsection{Case Studies}

Case studies provided insights into Bugdar’s performance by highlighting both successful detections and areas of improvement. Bugdar demonstrated its effectiveness in identifying vulnerabilities that were missed by traditional static analysis tools. For example, it \textbf{successfully detected a reentrancy vulnerability} in Solidity code, which other tools had overlooked. Another case showcased Bugdar’s ability to \textbf{identify insecure deserialization} in Python, providing developers with clear and actionable remediation steps to address the issue.


However, there were instances where Bugdar either missed or misclassified vulnerabilities, revealing areas for further enhancement. One notable example was its \textbf{failure to detect a logic flaw due to a limited understanding of domain-specific context}. Additionally, Bugdar incorrectly flagged the safe use of an "unsafe" block in Rust as a vulnerability, \textbf{demonstrating challenges in distinguishing between legitimate and problematic code usage}. These cases underscore the importance of improving Bugdar’s contextual understanding and refining its analysis capabilities for complex and specialized scenarios.







\subsection{User Studies}

Preliminary user studies with developers highlighted Bugdar's usability and influence on workflows, specifically, developers valued Bugdar’s concise, actionable commentary within pull requests. Moreover, developers consider early identification of issues can further reduce manual review time and streamline the development process.

\section{Discussion}

\subsection{Strengths, Challenges, and Limitation}

Bugdar demonstrates significant strengths in automating secure code reviews and providing actionable insights. By chunking code diffs based on the LLM's context window size, large code changes are efficiently processed without the overhead of AST parsing. However, further training may be required to detect complex, context-specific vulnerabilities. High false-positive rates in certain contexts can lead to alert fatigue among developers.

Moreover, the study identified several limitations that highlight areas for improvement and future development. One key limitation is the \textbf{variation in Bugdar’s effectiveness across different programming languages} and vulnerability types, underscoring the need for ongoing training to expand its coverage and accuracy. Additionally, \textbf{domain-specific code poses challenges}, as specialized implementations can lead to false positives, which may frustrate developers. Another issue is the \textbf{system’s reliance on training datasets}, which could introduce biases or leave gaps in its detection capabilities.






\subsection{Future Directions}

Future work on Bugdar focuses on several key areas for improvement and expansion. One priority is \textbf{enhancing the LLM's ability to understand domain-specific contexts} and complex business logic, which would improve its accuracy in identifying nuanced vulnerabilities. Another development area involves \textbf{exploring active learning techniques}, allowing Bugdar to continuously improve by incorporating developer feedback and adapting to new vulnerability patterns. Lastly, \textbf{integrating Bugdar more seamlessly into developer workflows} is a crucial goal, including the development of plugins for popular Integrated Development Environments (IDEs) and the creation of interactive dashboards \cite{chen2022decision} that provide real-time feedback and insights.



\section{Conclusion}

Bugdar represents a significant advancement in secure code review by leveraging AI to provide continuous, accurate, and project-specific security analysis within the GitHub workflow. The development of Bugdar is an ongoing journey, and continuous evaluation, user feedback, and research efforts will be essential for realizing its full potential and addressing the ever-evolving challenges of secure code review.

\section*{Acknowledgments}


\balance

\bibliographystyle{IEEEtran}
\bibliography{references}

\begin{thebibliography}{10}
\providecommand{\url}[1]{#1}
\csname url@samestyle\endcsname
\providecommand{\newblock}{\relax}
\providecommand{\bibinfo}[2]{#2}
\providecommand{\BIBentrySTDinterwordspacing}{\spaceskip=0pt\relax}
\providecommand{\BIBentryALTinterwordstretchfactor}{4}
\providecommand{\BIBentryALTinterwordspacing}{\spaceskip=\fontdimen2\font plus
\BIBentryALTinterwordstretchfactor\fontdimen3\font minus
  \fontdimen4\font\relax}
\providecommand{\BIBforeignlanguage}[2]{{%
\expandafter\ifx\csname l@#1\endcsname\relax
\typeout{** WARNING: IEEEtran.bst: No hyphenation pattern has been}%
\typeout{** loaded for the language `#1'. Using the pattern for}%
\typeout{** the default language instead.}%
\else
\language=\csname l@#1\endcsname
\fi
#2}}
\providecommand{\BIBdecl}{\relax}
\BIBdecl

\bibitem{cyberscope2022audit}
\BIBentryALTinterwordspacing
Cyberscope. (2022, Nov) How much does it cost to audit a smart contract? Medium
  Coinmonks. [Online]. Available:
  \url{https://medium.com/coinmonks/how-much-does-it-cost-to-audit-a-smart-contract-27ace328c0ce}
\BIBentrySTDinterwordspacing

\bibitem{6606613}
B.~Johnson, Y.~Song \emph{et~al.}, ``Why don't software developers use static
  analysis tools to find bugs?'' in \emph{2013 35th International Conference on
  Software Engineering (ICSE)}, 2013, pp. 672--681.

\bibitem{david2023manual}
\BIBentryALTinterwordspacing
I.~David \emph{et~al.}, ``Do you still need a manual smart contract audit?''
  \emph{arXiv preprint arXiv:2306.12338}, 2023. [Online]. Available:
  \url{https://arxiv.org/abs/2306.12338}
\BIBentrySTDinterwordspacing

\bibitem{gao2023far}
\BIBentryALTinterwordspacing
Z.~Gao, H.~Wang \emph{et~al.}, ``How far have we gone in vulnerability
  detection using large language models,'' \emph{arXiv preprint
  arXiv:2311.12420}, 2023. [Online]. Available:
  \url{https://arxiv.org/abs/2311.12420}
\BIBentrySTDinterwordspacing

\bibitem{li2021drawbacks}
\BIBentryALTinterwordspacing
Z.~Li, R.~Q. Shin \emph{et~al.}, ``Drawbacks in detecting vulnerabilities: A
  study of large language models on code,'' \emph{arXiv preprint
  arXiv:2105.00823}, 2021. [Online]. Available:
  \url{https://arxiv.org/abs/2105.00823}
\BIBentrySTDinterwordspacing

\bibitem{chen2024mad}
E.~Chen \emph{et~al.}, ``Suigpt mad: Move ai decompiler to improve transparency
  and auditability on non-open-source blockchain smart contract,'' \emph{arXiv
  preprint arXiv:2410.15275}, 2024.

\bibitem{chen2021evaluatinglargelanguagemodels}
\BIBentryALTinterwordspacing
M.~Chen \emph{et~al.}, ``Evaluating large language models trained on code,''
  2021. [Online]. Available: \url{https://arxiv.org/abs/2107.03374}
\BIBentrySTDinterwordspacing

\bibitem{chen2023gptutor}
E.~Chen, R.~Huang \emph{et~al.}, ``Gptutor: a chatgpt-powered programming tool
  for code explanation,'' in \emph{International Conference on Artificial
  Intelligence in Education}.\hskip 1em plus 0.5em minus 0.4em\relax Springer,
  2023, pp. 321--327.

\bibitem{achiam2023gpt}
J.~Achiam \emph{et~al.}, ``Gpt-4 technical report,'' \emph{arXiv preprint
  arXiv:2303.08774}, 2023.

\bibitem{li2018vuldeepecker}
\BIBentryALTinterwordspacing
Z.~Li, C.~Wang, S.~Zhu \emph{et~al.}, ``Vuldeepecker: A deep learning-based
  system for vulnerability detection,'' \emph{NDSS}, vol.~18, pp. 8--8, 2018.
  [Online]. Available:
  \url{https://www.ndss-symposium.org/wp-content/uploads/2018/02/ndss2018_03A-2_Li_paper.pdf}
\BIBentrySTDinterwordspacing

\bibitem{ragkhitwetsagul2019survey}
\BIBentryALTinterwordspacing
C.~Ragkhitwetsagul, S.~Boonphak, and C.~Tantithamthavorn, ``A survey on mining
  source code artifacts: Techniques, datasets, and tools,'' \emph{Journal of
  Systems and Software}, vol. 157, p. 110380, 2019. [Online]. Available:
  \url{https://doi.org/10.1016/j.jss.2019.110380}
\BIBentrySTDinterwordspacing

\bibitem{lewis2020retrievalaugmented}
\BIBentryALTinterwordspacing
P.~Lewis \emph{et~al.}, ``Retrieval-augmented generation for
  knowledge-intensive nlp tasks,'' \emph{arXiv preprint arXiv:2005.11401},
  2020. [Online]. Available: \url{https://arxiv.org/abs/2005.11401}
\BIBentrySTDinterwordspacing

\bibitem{blackshear2019move}
S.~Blackshear, E.~Cheng \emph{et~al.}, ``Move: A language with programmable
  resources,'' \emph{Libra Assoc}, vol.~1, 2019.

\bibitem{chen2022decision}
E.~Chen \emph{et~al.}, ``A decision model for designing nlp applications,'' in
  \emph{Companion Proceedings of the Web Conference}, 2022, pp. 1206--1210.

\end{thebibliography}

\end{document}